\documentclass[12pt,preprint]{aastex} 		

\usepackage{natbib}

\newcommand{\msun}{M$_{\sun}$}
\newcommand{\av}{$A_V$}
\newcommand{\etal}{et~al.}
\newcommand{\ic}{$I_{\rm C}$}
\newcommand{\ks}{$K_{\rm s}$}
\newcommand{\teff}{$T_{\rm eff}$}
\newcommand{\mum}{$\mu$m}

\begin{document}

\title{Near and Mid-IR Photometry of the Pleiades, and a New List
of Substellar Candidate Members\altaffilmark{1,2}}

\slugcomment{ApJS, in press; version with embedded figures can be
obtained at http://spider.ipac.caltech.edu/staff/stauffer/} 
\shortauthors{Stauffer et al.} 
\shorttitle{Infrared Observations of the Pleiades}

\author{John R.\ Stauffer}
\affil{Spitzer Science Center, Caltech 314-6, Pasadena, CA  91125}
\email{stauffer@ipac.caltech.edu}

\author{Lee W.\ Hartmann}
\affil{Astronomy Department, University of Michigan }

\author{Giovanni G.\ Fazio, Lori E.\ Allen, Brian M.\ Patten}
\affil{Harvard-Smithsonian Center for Astrophysics, 60 Garden St.,
Cambridge, MA  02138}

\author{Patrick J.\ Lowrance, Robert L.\ Hurt, Luisa M.\ Rebull}
\affil{Spitzer Science Center, Caltech , Pasadena, CA 91125}

\author{Roc M.\ Cutri, Solange V.\ Ramirez}
\affil{Infrared Processing and Analysis Center, Caltech 220-6, Pasadena, CA  91125}

\author{Erick T.\ Young, George H.\ Rieke, Nadya I.\ Gorlova\altaffilmark{3}, James C.\ Muzerolle}
\affil{Steward Observatory, University of Arizona, Tucson,
AZ  85726}

\author{Cathy L.\ Slesnick}
\affil{Astronomy Department, Caltech, Pasadena, CA 91125}

\author{Michael F.\ Skrutskie}
\affil{Astronomy Department, University of Virginia, Charlottesville, VA  22903}

\altaffiltext{1}{This work is based (in part) on observations made
with the Spitzer Space Telescope, which is operated by the Jet 
Propulsion Laboratory, California Institute of Technology, under
NASA contract 1407. } 
\altaffiltext{2}{This publication makes use of data products 
from the Two Micron All Sky Survey, which is a joint project 
of the University of Massachusetts and the Infrared Processing 
and Analysis Center/California Institute of Technology, 
funded by the National Aeronautics and Space Administration 
and the National Science Foundation.}
\altaffiltext{3}{Current address: University of Florida, 211 Bryant Space
Center, Gainesville, FL  32611}

\begin{abstract}

We  make use of new near and mid-IR photometry of the Pleiades cluster
in order to help identify proposed  cluster members.  We also use the
new photometry with previously published photometry to define the
single-star main sequence locus at the age of the Pleiades in a
variety of color-magnitude planes. 

The new near and mid-IR photometry extend effectively two magnitudes
deeper than the 2MASS All-Sky Point Source catalog, and hence allow us
to select a new set of candidate very low mass and  sub-stellar mass
members of the Pleiades in the central square degree of the cluster.
We identify 42 new candidate members fainter than Ks =14
(corresponding to 0.1 Mo). These candidate members should eventually
allow a better estimate of the cluster mass function to be made down
to of order 0.04 solar masses.

We also use new IRAC data, in particular the images obtained at  8 um,
in order to comment briefly on interstellar dust in and near the
Pleiades.  We confirm, as expected, that -- with one exception -- a
sample of low mass stars recently identified as having 24 um excesses
due to debris disks do not have significant excesses at IRAC
wavelengths.  However, evidence is also presented that  several of the
Pleiades high mass stars are found to be impacting with local
condensations of the molecular cloud that is passing through the
Pleiades at the current epoch.

\end{abstract}

\keywords{
stars: low mass ---
young; open clusters ---
associations: individual (Pleiades)
}

\section{Introduction}
\label{sec:intro}

Because of its proximity, youth, richness, and location in the
northern hemisphere, the Pleiades has long been a favorite target of
observers. The Pleiades was one of the first open clusters to have
members identified via their common proper motion \citep{trumpler21},
and the cluster has since then been the subject of more than a dozen
proper motion studies. Some of the earliest photoelectric photometry
was for members of the Pleiades \citep{cummings21}, and the cluster
has been the subject of dozens of papers providing additional optical
photometry of its members.  The youth and nearness of the Pleiades
make it a particularly attractive target for identifying its
substellar population, and it was the first open cluster studied for
those purposes \citep{jameson89,stauffer89}.  More than 20  papers
have been subsequently published, identifying additional substellar
candidate members of the Pleiades or studying their properties.

We have three primary goals for this paper.   First, while
extensive optical photometry for Pleiades members is available in the
literature, photometry in the near and mid-IR is relatively spotty. 
We will remedy this situation by using new 2MASS $JHK_s$ and Spitzer IRAC
photometry for a large number of Pleiades members.  We will use these
data to help identify cluster non-members and to define the
single-star locus in color-magnitude diagrams for stars of 100 Myr
age.  Second, we will use our new IR imaging photometry of the center
of the Pleiades to identify a new set of candidate substellar members
of the cluster, extending down to stars expected to have masses of
order 0.04 \msun.   Third, we will use the IRAC data to briefly
comment on the presence of circumstellar debris disks in the Pleiades
and the interaction of the Pleiades stars with the molecular cloud
that is currently passing through the cluster.

In order to make best use of the IR imaging data,
we will begin with a necessary digression.   As noted
above, more than a  dozen proper motion surveys of the Pleiades
have been made in order to identify cluster members.   However,
no single catalog of the cluster has been published which
attempts to collect all of those candidate members in a single
table and cross-identify those stars.  Another problem is that
while there have been many papers devoted to providing optical
photometry of cluster members, that photometry has been
bewilderingly inhomogeneous in terms of the number of
photometric systems used.   In Sec.\ 3 and in the Appendix,
we describe our efforts to create a  reasonably complete catalog
of candidate Pleiades members and to provide optical photometry
transformed to the best of our ability onto a
single system.

\section{New Observational Data}
\label{sec:observations}

\subsection{2MASS ``6x" Imaging of the Pleiades}

During the final months of Two Micron All Sky Survey (2MASS; \citet{skrutskie06})
operations, a series of special observations were carried out that
employed exposures six times longer than used for the the primary survey.
These so-called ``6x" observations targeted 30 regions of scientific interest
including a 3 deg  $x$ 2 deg area centered on the Pleiades cluster.  The 2MASS
6x data were reduced using an automated processing pipeline similar to that
used for the main survey data, and a calibrated 6x Image Atlas and extracted
6x Point and Extended Source Catalogs (6x-PSC and 6x-XSC) analogous to the
2MASS All-Sky Atlas, PSC and XSC have been released as part of the 2MASS
Extended Mission.  A description of the content and formats of the 6x image
and catalog products, and details about the 6x observations and data reduction
are given by Cutri et al. (2006; section A3).
\footnote{http://www.ipac.caltech.edu/2mass/releases/allsky/doc/explsup.html} 
The 2MASS 6x Atlas and Catalogs
may be accessed via the on-line services of the NASA/IPAC Infrared Science
Archive (http://irsa.ipac.caltech.edu).

Figure 1 shows the area on the sky imaged by the 2MASS 6x observations
in the Pleiades field.  The region was covered by two rows of scans, each
scan being one degree long (in declination) and 8.5' wide in right
ascension.  Within each row, the scans overlap by approximately one arcminute
in right ascension.  There are small gaps in coverage in the declination
boundary between the rows, and one complete scan in the southern row is
missing because the data in that scan did not meet the minimum required
photometric quality.   The total area covered by the 6x Pleiades observations
is approximately 5.3 sq. degrees.

There are approximately 43,000 sources extracted from the 6x Pleiades
observations in the 2MASS 6x-PSC, and nearly 1,500 in the 6x-XSC. Because
there are at most about 1000 Pleiades members expected in this region, only
$\sim$2\% of the 6x-PSC sources are cluster members, and the rest are field stars
and background galaxies.  The 6x-XSC objects are virtually all resolved
background galaxies.  Near infrared color-magnitude and color-color diagrams
of the unresolved sources from the 2MASS 6x-PSC and all sources in the 6x-XSC
sources from the Pleiades region are shown in Figures 2 and 3, respectively.
The extragalactic sources tend to be redder than most stars, and the galaxies
become relatively more numerous towards fainter magnitudes.  Unresolved
galaxies dominate the point sources that are fainter than $K_s$ $>$ 15.5 and redder
than $J-K_s >$ 1.2 mag.

The 2MASS 6x observations were conducted using the same freeze-frame scanning
technique used for the primary survey \citep{skrutskie06}.  The longer
exposure times were achieved by increasing the ``READ2-READ1" integration to
7.8 sec from the 1.3 sec used for primary survey.  However, the 51 ms ``READ1"
exposure time was not changed for the 6x observations.  As a result,
there is an effective ``sensitivity gap" in the 8-11 mag region where objects
may be saturated in the 7.8 sec READ2-READ1 6x exposures, but too faint to
be detected in the 51 ms READ1 exposures.  Because the sensitivity gap can
result in incompleteness and/or flux bias in the photometric overlap regime,
the near infrared photometry for sources brighter than J=11 mag in the 6x-PSC
was taken from the 2MASS All-Sky PSC during compilation of the catalog
of Pleiades candidate members presented in Table 2 (c.f. Section 3).

\subsection{Shallow IRAC Imaging}

Imaging of the Pleiades with Spitzer was obtained in April 2004
as part of a joint GTO program conducted by the IRAC instrument
team and the MIPS instrument team.  Initial results of the MIPS
survey of the Pleiades have already been reported in
\citet{gorlova06}.  The IRAC observations were obtained as two
astronomical observing requests (AORs).  One of them was
centered near the cluster center, at RA=03h47m00.0s and
Dec=24d07m (2000), and consisted of a 12 row by 12 column map,
with ``frametimes" of 0.6 and 12.0 seconds and two dithers at
each map position.   The map steps were 290$\arcsec$ in both
the column and row direction.  The resultant map covers a region
of approximately one square degree, and a total integration time
per position of 24 sec over most of the map.  The second AOR
used the same basic mapping parameters, except it was smaller (9
rows by 9 columns) and was instead centered northwest from the
cluster center at RA=03h44m36.0s and Dec=25d24m.  A two-band
color image of the AOR covering the center of the Pleiades is
shown in Figure~\ref{fig:pleIRAC}.  A pictorial guide to the
IRAC image providing Greek names for a few of the brightest
stars, and  \citet{hertzsprung47} numbers for several stars
mentioned in Section 6 is provided in Figure~\ref{fig:cartoon}.

We began our analysis with the basic calibrated data (BCDs) from
the Spitzer pipeline, using the S13 version of the Spitzer Science 
Center pipeline
software.  Artifact  mitigation and masking was done using the
IDL tools provided on the Spitzer contributed software website. 
For each AOR, the artifact-corrected BCDs were  combined into
single mosaics for each channel using the post-BCD ``MOPEX"
package \citep{makovoz05}. The mosaic images were constructed
with 1.22$\times$1.22 arcsecond pixels (i.e., approximately the
same pixel size as the native IRAC arrays).

We derived aperture photometry for stars present in these IRAC
mosaics using both APEX (a component of the MOPEX package) and
the ``phot" routine in DAOPHOT.  In both cases, we used a 3 pixel
radius aperture and a sky annulus from 3 to 7 pixels (except
that for Channel 4, for the phot package we used a 2 pixel
radius aperture and a 2 to 6 pixel annulus because that provided
more reliable fluxes at low flux levels).   We used the flux for
zero magnitude calibrations provided in the IRAC data handbook
(280.9, 179.7, 115.0 and 64.1 Jy for Ch 1 through Ch 4,
respectively), and the aperture corrections provided in the same
handbook (multiplicative flux correction factors of
1.124, 1.127, 1.143 and 1.584 for Ch 1-4, inclusive.  The Ch4
correction factor is much bigger because it is for an aperture
radius of 2 rather than 3 pixels.).   

Figure~\ref{fig:plecomp1} and Figure~\ref{fig:plecomp2} provide
two means to assess the accuracy of the IRAC photometry.   The
first figure compares the aperture photometry from APEX to that
from phot, and shows that the two packages yield very similar
results when used in the same way.  For this reason, we have
simply averaged the fluxes from the two packages to obtain our
final reported value.  The second figure shows the difference
between the derived 3.6 and 4.5 \mum\ magnitudes for Pleiades
members.  Based on previous studies (e.g. \citet{allen04}), 
we expected this difference
to be essentially zero for most stars, and the Pleiades data
corroborate that expectation.  For [3.6]$<$10.5, the RMS
dispersion of the magnitude difference between the two channels
is 0.024 mag.  Assuming that each channel has similar
uncertainties, this indicates an internal 1-$\sigma$ accuracy of
order 0.017 mag.  The absolute calibration uncertainty for the
IRAC fluxes is currently estimated at of order 0.02 mag.  
Figure~\ref{fig:plecomp2} also shows that fainter than [3.6]=10.5
(spectral type later than about M0), the [3.6]$-$[4.5] color for
M dwarfs departs slightly from zero, becoming increasingly redder
to the limit of the data (about M6).

\section{A Catalog of Pleiades Candidate Members} 
\label{sec:catalog}

If one limits oneself to only stars visible with the naked eye,
it is easy to identify which stars are members of the Pleiades --
all of the stars within a degree of the cluster center that have
$V<$ 6 are indeed members.   However, if one were to try to
identify the M dwarf stellar members of the cluster (roughly 14
$<V<$ 23), only of order 1\% of the stars towards the
cluster center are likely to be members, and it is much harder
to construct an uncontaminated catalog.  The problem is
exacerbated by the fact that the Pleiades is old enough that
mass segregation through dynamical processes has occurred, and
therefore one has to survey a much larger region of the sky in
order to include  all of the M dwarf members.

The other primary difficulty in constructing a comprehensive
member catalog for the Pleiades is that the pedigree of the
candidates varies greatly.   For the best studied stars,
astrometric positions can be measured over temporal baselines
ranging up to a century or more, and the separation of cluster
members from field stars in a vector point diagram (VPD) can be
extremely good.   In addition, accurate radial velocities and
other spectral indicators are available for essentially all of
the  bright cluster members, and these further allow membership
assessment to be essentially  definitive.  Conversely, at the
faint end (for stars near the hydrogen burning mass limit in the
Pleiades), members are near the detection limit of the existing
wide-field photographic plates, and the errors on the proper
motions become correspondingly large, causing the separation of
cluster members from field stars in the VPD to become poor. 
These stars are also sufficiently faint that spectra capable
of discriminating members from field dwarfs can only be
obtained with 8m class telescopes, and only a very small
fraction of the faint candidates have had such spectra obtained.
Therefore, any comprehensive catalog created for the Pleiades
will necessarily have stars ranging from certain members to
candidates for which very little is known, and where the
fraction of spurious candidate members increases to lower
masses.

In order to address the membership uncertainties and biases, we
have chosen a sliding scale for inclusion in our
catalog. For all stars, we require that the available photometry
yields location in color-color and color-magnitude diagrams
consistent with cluster membership.  For the
stars with well-calibrated photoelectric photometry, this means
the star should not fall below the Pleiades single-star locus by
more than about 0.2 mag or above that locus by more than
about 1.0 mag (the expected displacement for a
hierarchical triple with three nearly equal mass components). 
For stars with only photographic optical photometry, where the
1-$\sigma$ uncertainties are of order 0.1 to 0.2 mag, we
still require the star's photometry to be consistent with
membership, but the allowed displacements from the single star
locus are considerably larger.   Where accurate radial
velocities are known, we require that the star be considered a
radial velocity member based on the paper where the radial
velocities were presented.    Where stars have been previously
identified as non-members based on photometric or spectroscopic
indices, we adopt those conclusions.

Two other relevant pieces of information are sometimes available. 
In some cases, individual proper motion membership probabilities
are provided by the various membership surveys.  If no other
information is available, and if the membership probability for
a given candidate is less than 0.1, we exclude that star from
our final catalog.  However, often a star appears in several
catalogs; if it appears in two or more proper motion
membership lists we include it in the final catalog even if P
$<$ 0.1 in one of those catalogs.   Second, an entirely
different means to identify candidate Pleiades members is via
flare star surveys towards the cluster \citep{haro82,jones81}.  
A star with a formally low membership probability in one catalog
but whose photometry is consistent with membership and that was
identified as a flare star is retained in our catalog.

Further details of the catalog construction are provided in the
appendix, as are details of the means by which the $B$, $V$, and
$I$ photometry have been homogenized.  A full discussion and listing
of all of the papers from which we have extracted astrometric and
photometric information is also provided in the appendix.   
Here we simply provide a
very brief description of the inputs to the catalog.

We include candidate cluster members from the following proper
motion surveys:  \citet{trumpler21}, \citet{hertzsprung47},
\citet{jones81}, Pels and Lub -- as reported in
\citet{vanlee86},  \citet{stauffer91}, \citet{artyukhina69},
\citet{hambly93}, \citet{pinfield00}, \citet{adams01} and \citet{deacon04}.   
Another important compilation which provides the initial
identification of a significant number of low mass cluster
members is the flare star catalog of \citet{haro82}. Table 1
provides a  brief synopsis of the characteristics of the
candidate member catalogs from these papers.  The Trumpler paper
is listed twice in Table 1 because there are two membership
surveys included in that paper, with differing spatial coverages
and different limiting magnitudes.

In our final catalog, we have attempted to follow the standard
naming convention whereby the primary name 
is derived from the paper where it was first identified
as a cluster member.  An exception to this arises for stars with
both \citet{trumpler21} and \citet{hertzsprung47} names, where
we use the Hertzsprung numbers as the standard name because that
is the most commonly used designation for these stars in the
literature.   The failure for the Trumpler numbers to be given
precedence in the literature perhaps stems from the fact that
the Trumpler catalog was published in the Lick Observatory
Bulletins as opposed to a refereed journal. In addition to
providing a primary name for each star, we provide
cross-identifications to some of the other catalogs,
particularly where there is existing photometry or spectroscopy
of that star using the alternate names.  For the brightest
cluster members, we provide additional cross-references (e.g.,
Greek names, Flamsteed numbers, HD numbers).

For each star, we attempt to include an estimate for Johnson $B$
and $V$, and for Cousins $I$ (\ic).   Only a very small fraction of
the cluster members have photoelectric photometry in these
systems, unfortunately. Photometry for many of the stars has
often been obtained in other systems, including Walraven,
Geneva, Kron, and Johnson.  We have used previously published
transformations from the appropriate indices in those systems to
Johnson $BV$ or Cousins $I$.   In other cases, photometry is
available in a natural $I$ band system, primarily for some of the
relatively faint cluster members. We have attempted to transform
those $I$ band data to \ic\ by deriving our own
conversion using stars for which we already have a \ic\ estimate
as well as the natural $I$ measurement.  Details of these issues
are provided in the Appendix.

Finally, we have cross-correlated the cluster candidates catalog
with the 2MASS All-Sky PSC and also with the 6x-PSC
for the Pleiades.   For every star in the catalog, we 
obtain $JH$\ks\ photometry and 2MASS positions.  Where we have
both main survey 2MASS data and data from the 6x catalog, we
adopt the 6x data for stars with $J>$11, and data from the
standard  2MASS catalog otherwise.   We verified that
the two catalogs do not have any obvious photometric or
astrometric offsets relative to each other.    The
coordinates we list in our catalog are entirely from these 2MASS
sources, and hence they inherit the very good and homogeneous
2MASS positional accuracies of order 0.1 arcseconds RMS.

We have then plotted the candidate Pleiades members in a variety
of color-magnitude diagrams and color-color diagrams, and
required that a star must have photometry that is consistent
with cluster membership.   Figure~\ref{fig:ple1695} illustrates
this process, and indicates why (for example) we have excluded
HII 1695  from our final catalog.

Table 2 provides the collected data for the 1417 stars we have
retained as candidate Pleiades members.  The first two columns are the
J2000 RA and Dec from 2MASS; the next are the 2MASS $JH$\ks\ photometry and
their uncertainties, and the 2MASS photometric quality flag (``ph-qual").  
If the number
following the 2MASS quality flag is a 1, the 2MASS data come from
the 2MASS All-Sky PSC; if it is a 2, the data come from the 6x-PSC.
The next three columns provide the $B$, $V$ and \ic\
photometry, followed by a flag which indicates the provenance of that
photometry.   The last column provides the most commonly used names
for these stars.   The hydrogen burning mass limit for the Pleiades
occurs at about $V$=22, $I$=18, \ks=14.4.   Fifty-three of the
candidate members in the catalog are fainter than this limit, and
hence should be sub-stellar if they are indeed Pleiades members.

Table 3 provides the IRAC [3.6], [4.5], [5.8] and [8.0]  photometry we
have derived for Pleiades candidate members included within the region
covered by the IRAC shallow survey of the Pleiades (see section 2). 
The brightest stars are saturated even in our short integration frame
data, particularly for the more sensitive 3.6 and 4.5 \mum\ channels. 
At the faint end, we provide photometry only for 3.6 and 4.5 \mum\
because the objects are undetected in the two longer wavelength
channels.  At the ``top" and ``bottom" of the survey region, we have
incomplete wavelength coverage for a band of width about 5$\arcmin$,
and for stars in those areas we report only photometry in either the
3.6 and 5.8 bands or in 4.5 and 8.0 bands.  

Because Table 2 is an amalgam of many previous catalogs, each of
which have different spatial coverage, magnitude limits and
other idiosyncrasies, it is necessarily incomplete and
inhomogeneous. It also certainly includes some non-members.  For
$V<$ 12, we expect very few non-members because of the extensive
spectroscopic data available for those stars; the fraction of
non-members will likely increase to fainter magnitudes,
particularly for stars located far from the cluster center. The
catalog is simply an attempt to collect all of the available
data, identify some of the non-members and eliminate
duplications. We hope that it will also serve as a starting
point for future efforts to produce a ``cleaner" catalog.

Figure~\ref{fig:plespatial2} shows the distribution on the sky
of the stars in Table 2. The complete spatial distribution of
all members of the Pleiades may differ slightly from what is
shown due to the inhomogeneous properties of the proper motion
surveys.  However, we believe that those effects are relatively
small and the distribution shown is mostly representative of the
parent population.  One  thing that is evident in  Figure
\ref{fig:plespatial2} is mass segregation -- the highest mass
cluster members are much more centrally located than the lowest
mass cluster members.   This fact is reinforced by calculating
the cumulative number of stars as a function of distance from
the cluster center for different absolute magnitude bins. 
Figure~\ref{fig:ple_segreg}  illustrates this fact.  Another
property of the Pleiades illustrated by Figure
\ref{fig:plespatial2} is that the cluster appears to be
elongated parallel to the galactic plane, as expected from n-body
simulations of galactic clusters \citep{terlevich87}.  Similar
plots showing the flattening of the cluster and evidence for
mass segregation for the V $<$ 12 cluster members were provided
by \citep{raboud98}.

\section{Empirical Pleiades Isochrones and Comparison to Model Isochrones}

Young, nearby, rich open clusters like the Pleiades can and
should be used to provide template data which can 
help interpret observations of more distant clusters or to 
test theoretical models.   The identification of candidate
members of distant open clusters is often based on plots of
stars in a color-magnitude diagram, overlaid upon which is a
line meant to define the single-star locus at the distance of
the cluster.   The stars lying near or slightly above the locus
are chosen as possible or probable cluster members.  The data we
have collected for the Pleiades provide a means to define the
single-star locus for  100 Myr, solar metallicity stars in a
variety of widely used color systems down to and slightly below
the hydrogen burning mass limit.  Figure~\ref{fig:cmd_vmi} and
Figure~\ref{fig:cmd_km1} illustrate the appearance of the
Pleiades stars in two of these diagrams, and the single-star
locus we have defined.   The curve defining the single-star
locus was drawn entirely ``by eye.''  It is displaced slightly
above the lower envelope to the locus of stars to
account for photometric uncertainties (which increase to fainter
magnitudes).  We attempted to use all of the information
available to us, however.   That is, there should also be an
upper envelope to the Pleiades locus in these diagrams, since
equal mass binaries should be displaced above the single star
sequence by 0.7 magnitudes (and one expects very few systems of
higher multiplicity).   Therefore, the single star locus was
defined with that upper envelope in mind.  Table 4 provides the
single-star loci for the Pleiades for $BVI_{\rm c}JK_{\rm s}$
plus the four IRAC channels.   We have  dereddened the
empirical loci by the canonical mean extinction to the Pleiades
of \av\ = 0.12 (and, correspondingly, A$_B$ = 0.16, A$_I$ =
0.07, A$_J$ = 0.03, A$_K$ = 0.01, as per the reddening law
of \citet{rieke85}).

The other benefit to constructing the new catalog is that it can
provide an improved comparison dataset to test theoretical
isochrones.  The new catalog provides homogeneous photometry in
many photometric bands  for stars ranging from several solar
masses down to below 0.1 \msun. 
We take the distance to the Pleiades
as 133 pc, and refer the reader to \citet{soderblom05} for a
discussion and a listing of the most recent determinations.  The
age of the Pleiades is not as well-defined, but is probably
somewhere between 100 and 125 Myr \citep{meynet93, stauffer98}. 
We adopt 100 Myr for the purposes of this discussion; our
conclusions relative to the theoretical isochrones would not be
affected significantly if we instead chose 125 Myr.  As noted
above, we adopt \av=0.12 as the mean Pleiades extinction, and
apply that value to the theoretical isochrones. A small number
of Pleiades members have significantly larger extinctions
\citep{breger86, stauffer87}, and we have dereddened those
stars individually to the mean cluster reddening.

Figures \ref{fig:super_vmi} and \ref{fig:super_kik} compare
theoretical 100 Myr isochrones from \citet{siess00} and
\citet{baraffe98} to the Pleiades member photometry from Table 2
for stars for which we have photoelectric photometry.  Neither
set of isochrones are a good fit to the $V-I$ based
color-magnitude diagram.  For \citet{baraffe98} this is not a
surprise because they illustrated that their isochrones are too
blue in $V-I$ for cool stars in their paper, and ascribed the
problem as likely the result of an incomplete line list, 
resulting in too little absorption in the $V$ band.  For
\citet{siess00}, the poor fit in the $V-I$ CMD is somewhat
unexpected in that they transform from the theoretical to the
observational plane using empirical color-temperature
relations.  In any event, it is clear that neither model
isochrones  match the shape of the Pleiades locus in the $V$ vs.\
$V-I$ plane, and therefore use of these $V-I$ based isochrones for
younger clusters is not likely to yield accurate results (unless
the color-\teff\ relation is recalibrated, as  described for
example in \citet{jeffries05}).  On the other hand,  the
\citet{baraffe98} model provides a quite good fit to the
Pleiades single star locus for an age of 100 Myr in the $K$ vs.\
$I-K$ plane.\footnote{These isochrones are calculated for the
standard K filter, rather than \ks.   However, the difference in
location of the isochrones in these plots because of this should
be very slight, and we do not believe our conclusions are significantly
affected.}.
This perhaps lends support to the hypothesis that
the misfit in the $V$ vs.\ $V-I$ plane is due to missing opacity in
their V band atmospheres for low mass stars (see also \citet{chabrier00}
for further evidence in support of this idea).  The
\citet{siess00} isochrones do not fit the Pleiades locus in the
$K$ vs.\ $I-K$ plane particularly well, being too faint near
$I-K$=2 and too bright for $I-K >$ 2.5.  

\section{Identification of New Very Low Mass Candidate  Members}

The highest spatial density for Pleiades members of any mass
should be at the cluster center.   However, searches for
substellar members of the Pleiades have generally avoided the
cluster center because of the deleterious effects of scattered
light from the high mass cluster members and because of the
variable background from the Pleiades reflection nebulae.   The
deep 2MASS and IRAC 3.6 and 4.5 \mum\ imaging provide accurate
photometry to well below the hydrogen burning mass limit, and
are less affected by the nebular emission than shorter
wavelength images.   We therefore expect that it should be
possible to identify a new set of candidate Pleiades substellar
members by combining our new near and mid-infrared photometry.

The  substellar mass limit in the Pleiades occurs at about
\ks =14.4, near the limit of the 2MASS All-Sky PSC.  As
illustrated in Figure  \ref{fig:2macmd}, the deep 2MASS survey
of the Pleiades should easily detect objects at least two
magnitudes fainter than the substellar limit.  The key to
actually identifying those objects and separating them from the
background sources is to find color-magnitude or color-color
diagrams  which separate the Pleiades members from the other
objects. As shown in Figure~\ref{fig:cmd3dot6}, late-type
Pleiades members separate fairly well from most field stars
towards the Pleiades in a \ks\ vs.\ $K_s-[3.6]$ color-magnitude
diagram.  However, as illustrated in Figure~\ref{fig:2macmd}, in
the $K_s$ magnitude range of interest there is also a large
population of red galaxies, and they are in fact the primary
contaminants to identifying Pleiades substellar objects in the \ks\ 
vs.\ $K_s-[3.6]$ plane.  Fortunately, most of the contaminant
galaxies are slightly resolved in the 2MASS and IRAC imaging,
and we have found that we can eliminate most of the red galaxies
by their non-stellar image shape.

Figure~\ref{fig:cmd3dot6} shows the first step in our process of
identifying new very low mass members of the Pleiades.   The
red plus symbols are the known Pleiades members from Table 2. 
The red open circles are candidate Pleiades substellar members
from deep imaging surveys published in the literature, mostly of
parts of the cluster exterior to the central square degree,
where the IRAC photometry is from \citet{lowrance07}.  The blue,
filled circles are field M and L dwarfs, placed at the distance
of the Pleiades, using photometry from \citet{patten06}. 
Because the Pleiades is $\sim$100 Myr, its very low mass stellar
and substellar objects will be displaced about 0.7 mag above the
locus of the field M and L dwarfs according to the
\citet{baraffe98} and \citet{chabrier00} models, in accord with the location in the
diagram of the previously identified, candidate VLM and
substellar objects.  The trapezoidal shaped region outlined with
a dashed line is the region in the diagram which we define as
containing candidate new VLM and substellar members of the
Pleiades.   We place the faint limit of this region at \ks =16.2
in order to avoid the large apparent increase in faint, red
objects for \ks $>$ 16.2, caused largely by increasing errors in
the \ks\ photometry.  Also, the 2MASS extended object flags cease
to be useful fainter than about \ks= 16.

We took the following steps to identify a set of candidate substellar
members of the Pleiades:
\begin{itemize}
\item keep only objects which fall in the trapezoidal region in
Figure~\ref{fig:cmd3dot6}.
\item  remove objects flagged as non-stellar by the 2MASS pipeline software;
\item  remove objects which appear non-stellar to the eye in the IRAC images;
\item  remove objects which do not fall in or near the locus of
field M and L dwarfs in a $J-H$ vs.\ $H-K_s$  diagram;
\item  remove objects which have 3.6 and 4.5 \mum\ magnitudes that differ
by more than 0.2 mag.
\item  remove objects which fall below the ZAMS in a J vs. $J-K_s$ diagram.
\end{itemize}
As shown in Figure~\ref{fig:cmd3dot6}, all stars earlier than
about mid-M have  $K_s-[3.6]$ colors bluer than 0.4.  This ensures
that for most of the area of the trapezoidal region, the primary
contaminants are distant galaxies.  Fortunately, the 2MASS
catalog provides two types of flags for identifying extended
objects. For each filter, a chi-square flag measures the match
between the objects shape and the instrumental PSF, with values
greater than 2.0 generally indicative of a non-stellar object. 
In order not to be misguided by an image artifact in one filter,
we throw out the most discrepant of the three flags and average
the other two.  We discard objects with mean $\chi^2$ greater
than 1.9.  The other indicator is the 2MASS extended object
flag, which is the synthesis of several independent tests of the
objects shape, surface brightness and color (see \citet{jarrett00}
for a description of this process). If one
simply excludes the objects classified as extended in the 2MASS
6x image by either of these techniques, the number of
candidate VLM and  substellar objects lying inside the
trapezoidal region decreases by nearly a half.

We have one additional means to demonstrate that many
of the identified objects are probably Pleiades members, and
that is via proper motions.  The mean Pleiades proper motion is
$\Delta$RA = 20 mas yr$^{-1}$ and $\Delta$Dec = $-$45 mas
yr$^{-1}$ \citep{jones73}.  With an epoch difference of only 3.5
years between the deep 2MASS and IRAC imaging, the expected
motion for a Pleiades member is only 0.07 arcseconds in RA and
$-$0.16 arcseconds in Dec.  Given the relatively large pixel
size for the two cameras, and the undersampled nature of the
IRAC 3.6 and 4.5 \mum\ images, it is not a priori obvious that
one would expect to reliably detect the Pleiades motion. 
However, both the 2MASS and IRAC astrometric solutions have been
very accurately calibrated. Also, for the present purpose, we
only ask whether the data support a conclusion that most of the
identified substellar candidates are true Pleiades members
(i.e., as an ensemble), rather than that each star is well
enough separated in a VPD to derive a high membership
probability.

Figure~\ref{fig:super_propmo} provides a set of plots that we
believe support the conclusion that the majority of the
surviving VLM and substellar candidates  are Pleiades members. 
The first plot shows the measured motions between  the epoch of
the 2MASS and IRAC observations for all known Pleiades members
from Table 2 that lie in the central square degree region and
have 11 $<$ \ks\ $<$ 14 (i.e., just brighter than the substellar
candidates). The mean offset of the Pleiades stellar members
from the background population is well-defined and is
quantitatively of the expected magnitude and sign (+0.07 arcsec
in RA and $-$0.16 arcsec in Dec).  The RMS dispersion of the
coordinate difference for the  field population in RA and Dec is
0.076 and 0.062 arcseconds, supportive of our claim that the
relative astrometry for the two cameras is quite good.  Because
we expect that the background population should have essentially
no mean proper motion, the non-zero mean ``motion" of the field
population of about $<\Delta$RA$>$=0.3 arcseconds is
presumably not real.   Instead, the offset is probably due to
the uncertainty in transferring the Spitzer coordinate
zero-point between the warm star-tracker and the cryogenic focal
plane.  Because it is simply a zero-point offset applicable to
all the objects in the IRAC catalog, it has no effect on the
ability to separate Pleiades members from the field star
population.

The second panel in Figure~\ref{fig:super_propmo} shows the
proper motion of the candidate Pleiades VLM and substellar
objects.  While these objects do not show as clean a
distribution as the known members, their mean motion is clearly
in the same direction.   After removing 2-$\sigma$ deviants, the
median offsets for the substellar candidates are 0.04 and
$-$0.11 arcseconds in RA and Dec, respectively. The objects
whose motions differ significantly from the Pleiades mean may be
non-members or they may be members with poorly determined
motions (since a few of the high probability members in the
first panel also show discrepant motions).

The other two panels in Figure~\ref{fig:super_propmo} show the proper
motions of two possible control samples.  The first control sample was
defined as the set of stars that fall up to 0.3 magnitudes below the
lower sloping boundary of  the trapezoid in
Figure~\ref{fig:cmd3dot6}.  These objects should be late type dwarfs
that are either older or more distant than the Pleiades or red galaxies. 
We used the 2MASS data to remove extended or blended objects from the
sample in the same way as for the Pleiades candidates.  If the
objects are nearby field stars, we expect to see large proper motions;
if galaxies, the real proper motions would be small -- but relatively
large apparent proper motions due  to poor centroiding or different
centroids at different effective wavelengths could be present.  The
second control set was defined to have $-0.1 < K - [3.6] < 0.1$ and
$14.0 < K < 14.5$, and to be stellar based on the 2MASS flags.  This
control sample should therefore be relatively distant G and K dwarfs
primarily.   Both control samples have proper motion distributions
that differ greatly from the Pleiades samples and that make sense for,
respectively, a nearby and a distant field star sample.

Figure~\ref{fig:cmd3dot6memb} shows the Pleiades members from
Table 2 and the 55 candidate VLM and substellar members that
survived all of our culling steps. We cross-correlated this list
with the stars from Table 2 and with a list of the previously
identified candidate substellar members of the cluster from
other deep imaging surveys. Fourteen of the surviving objects
correspond to previously identified Pleiades VLM and substellar
candidates.   We provide the new list of candidate members in
Table 5.  The columns marked as $\mu$(RA) and $\mu$(DEC) are
the measured motions, in arcsec over the 3.5 year epoch difference
between the 2MASS-6x and IRAC observations.  
Forty-two of these objects have \ks $>$ 14.0, and hence
inferred masses less than about 0.1 \msun; thirty-one of them
have \ks $>$ 14.4, and hence have inferred masses below the
hydrogen burning mass limit.

Our candidate list could be contaminated by foreground late type
dwarfs that happen to lie in the line of sight to the Pleiades.  How
many such objects should we expect?  In order to pass our culling
steps, such stars would have to be mid to late M dwarfs, or early to
mid L dwarfs.  We use the known M dwarfs within 8 pc to estimate how
many field M dwarfs should lie in a one square degree region and at
distance between 70 and 100 parsecs (so they would be coincident in a
CMD with the 100 Myr Pleiades members). The result is $\sim$3 such
field M dwarf contaminants.  \citet{cruz06} estimate that the volume
density of L dwarfs is comparable to that for late-M dwarfs, and 
therefore a very conservative estimate is that there might also be 3
field L dwarfs contaminating our sample.   We regard this (6
contaminating field dwarfs) as an upper limit because our various
selection criteria would exclude early M dwarfs and late L dwarfs. 
\citet{bihain06} made an estimate of the number of contaminating
field dwarfs in their Pleiades survey of 1.8 square degrees; for the
spectral type range of our objects, their algorithm would have
predicted just one or two contaminating field dwarfs for our survey.

How many substellar Pleiades members should there be in the
region we have surveyed?  That is, of course, part of the
question we are trying to answer.  However, previous studies
have estimated that the Pleiades stellar mass function for M $<$
0.5 \msun\ can be approximated as a power-law with an exponent
of -1 (dN/dM $\propto$ M$^{-1}$).  Using the known Pleiades
members from Table 2 that lie within the region of the IRAC
survey and that have masses of 0.2 $<$ M/\msun $<$ 0.5 (as
estimated from the \citet{baraffe98} 100 Myr isochrone) to
normalize the relation, the M$^{-1}$\ mass function predicts
about 48 members in our search region and with 14 $<$ K $<$ 16.2
(corresponding to 0.1 $<$ M/\msun $<$ 0.035).  Other studies
have suggested that the mass function in the Pleiades becomes
shallower below 0.1 \msun, dN/dM $\propto$ M$^{-0.6}$.  Using
the same normalization as above, this functional form for the
Pleiades mass function for M $<$ 0.1 \msun\ yields a prediction
of 20 VLM and substellar members in our survey.   The number of
candidates we have found falls between these two estimates.  
Better proper motions and low-resolution spectroscopy will almost
certaintly eliminate some of these candidates as non-members.

\section{Mid-IR Observations of Dust and PAHS in the Pleiades}
\label{sec:discussion}

Since the earliest days of astrophotography, it has been clear
that the Pleiades stars are in relatively close proximity to
interstellar matter whose optical manifestation is the
spider-web like network of  filaments seen particularly strongly
towards several of the B stars in the cluster.   High resolution
spectra of the brightest Pleiades stars as well as CO maps
towards the cluster show that there is gas as well as dust
present, and that the (primary) interstellar cloud has a
significant radial velocity offset relative to the Pleiades
\citep{white03, federman84}. The gas and dust, therefore, are
not a remnant from the formation of the cluster but are simply
evidence of a a transitory event as this small cloud passes by
the cluster in our line of sight (see also \citet{breger86}).  
There are at least two claimed morphological signatures of a
direct interaction of the Pleiades with the cloud.  
\citet{white93} provided evidence that the IRAS 60 and 100 \mum\
image of the vicinity of the Pleiades showed a dark channel
immediately to the east of the Pleiades, which they interpreted
as the ``wake" of the Pleiades as it plowed through the cloud
from the east.  \citet{herbig01} provided a detailed analysis of
the optically brightest nebular feature in the Pleiades -- IC
349 (Barnard's Merope nebula) -- and concluded that the shape
and structure of that nebula could best be understood if the
cloud was running into the Pleiades from the southeast. 
\citet{herbig01} concluded that the IC 349 cloudlet, and by
extension the rest of the gas and dust enveloping the Pleiades,
are relatively distant outliers of the Taurus molecular clouds
(see also \citet{eggen50} for a much earlier discussion ascribing
the Merope nebulae as outliers of the Taurus clouds).
\citet{white03} has more recently proposed a hybrid model, where
there are two separate interstellar cloud complexes with very
different space motions, both of which are colliding
simultaneously with the Pleiades and with each other.

\citet{breger86} provided polarization measurements for a sample
of member and background stars towards the Pleiades, and argued
that the variation in polarization signatures across the face of
the cluster was evidence that some of the gas and dust was
within the cluster. In particular, Figure 6 of that paper showed
a fairly distinct interface region, with little residual
polarization to the NE portion of the cluster and an L-shaped
boundary running EW along the southern edge of the cluster and
then north-south along the western edge of the cluster.  Stars
to the south and west of that boundary show relatively large
polarizations and consistent angles (see also our Figure
\ref{fig:cartoon} where we provide a few polarization vectors
from \citet{breger86} to illustrate the location of the
interface region and the fact that the position angle of the
polarization correlates well with the location in the
interface).

There is a general correspondence between the polarization map
and what is seen with IRAC, in the sense that the B stars in the
NE portion of the cluster (Atlas and Alcyone)  have little
nebular emission in their vicinity, whereas those in the 
western part of the cluster (Maia, Electra and Asterope) have
prominent, filamentary dust emission in their vicinity.  The
L-shaped boundary is in fact visible in Figure~\ref{fig:pleIRAC}
as enhanced nebular emission running between and below a line
roughly joining Merope and Electra, and then making a right
angle and running  roughly parallel to a line running from
Electra to Maia to HII1234 (see Figure~\ref{fig:cartoon}).

\subsection{Pleiades Dust-Star Encounters Imaged with IRAC}
\label{sec: dust structures}

The Pleiades dust filaments are most strongly evident in IRAC's
8 \mum\ channel, as evidenced by the distinct red color of the
nebular features in Figure~\ref{fig:pleIRAC}.  The dominance at
8 \mum\ is  an expected feature of reflection nebulae, as
exemplified by NGC 7023 \citep{werner04}, where most of the
mid-infrared emission arises from polycyclic aromatic
hydrocarbons (PAHs) whose strongest bands in the 3 to 10 \mum\
region fall at 7.7 and 8.6 \mum. One might expect that if
portions of the passing cloud were particularly near to one of
the Pleiades members, it might be possible to identify such
interactions by searching for stars with 8.0 \mum\ excesses or
for stars with extended emission at 8 \mum. Figure
\ref{fig:dusty1} provides two such plots. Four stars stand out
as having significant extended 8 \mum\ emission, with two of
those stars also having an 8 \mum\ excess based on their
[3.6]$-$[8.0] color.   All of these stars, plus IC 349, are
located approximately along the interface region identified by
\citet{breger86}.

We have subtracted a PSF from the 8 \mum\ images for the stars
with extended emission, and those PSF-subtracted images are
provided in Figure~\ref{fig:psfsub}.   The image for HII 1234 has
the appearance of  a bow-shock.   The shape is reminiscent of
predictions for what one should expect from a collision between
a large cloud or a sheet of gas and an A star as described in
\citet{artymowicz97}. The \citet{artymowicz97} model posits that
A stars encountering a cloud will carve a paraboloidal shaped
cavity in the cloud via radiation pressure.  The exact size and
shape of the cavity depend on the relative velocity of the
encounter, the star's mass and luminosity and properties of the
ISM grains.   For typical parameters, the predicted
characteristic size of the cavity is of order 1000 AU, 
quite comparable to the size of the structures around HII 652 and
HII 1234. The observed appearance of the cavity depends on the
view angle to the  observer.  However, in any case, the
direction from which the gas is moving relative to the star can
be inferred from the location of the star relative to the curved
rim of the cavity; the ``wind" originates  approximately from
the direction connecting the star and the apex of the rim.   For
HII 1234, this indicates the cloud which it is encountering has a
motion relative to HII 1234 from the SSE, in accord with a
Taurus origin and not in accord for where a cloud is impacting
the Pleiades from the west as posited in \citet{white03}. The
nebular emission for HII 652 is less strongly bow-shaped, but the
peak of the excess emission is displaced roughly southward
from the star, consistent with the Taurus model and
inconsistent with gas flowing from the west.

Despite being the brightest part of the Pleiades nebulae in the
optical, IC 349 appears to be undetected in the  8 \mum\ image.
This is not because the 8 \mum\ image is insensitive to the
nebular emission - there is generally good agreement between
the structures seen in the optical and at 8 \mum, and most
of the filaments present in optical images of the Pleiades
are also visible on the 8 \mum\ image (see Figures
\ref{fig:pleIRAC} and \ref{fig:psfsub}) and even the psf-subtracted
image of Merope shows well-defined nebular filaments.  
The lack of enhanced 8 \mum\
emission from the region of IC 349 is probably 
because all of the small particles have been scoured away from this cloudlet,
consistent with Herbig's model to explain the HST surface
photometry and colors.   There is no PAH emission from
IC 349 because there are none of the small molecules that are the
postulated source of the PAH emission.  

IC349 is very bright in the optical, and undetected to a good sensitivity
limit at 8 \mum; it must be detectable via imaging at some wavelength
between 5000 \AA\ and 8 \mum.   We checked our 3.6 \mum\ data for
this purpose.  In the standard BCD mosaic image, we were unable to
discern an excess at the location of IC349 either simply by displaying
the image with various stretches or by doing cuts through the image.
We performed a PSF subtraction of Merope from the image in order to
attempt to improve our ability to detect faint, extended emission
30" from Merope - unfortunately, bright stars have ghost images 
in IRAC Ch. 1, and in this case the ghost image falls almost 
exactly at the location of IC349.  IC349 is also not detected in
visual inspection of our 2MASS 6x images.

\subsection{Circumstellar Disks and IRAC}

As part of the Spitzer FEPS (Formation and Evolution of Planetary
Systems) Legacy program,
using pointed MIPS photometry, \citet{stauffer05} identified
three G dwarfs in the Pleiades as having 24 \mum\ excesses
probably indicative of circumstellar dust disks.
\citet{gorlova06} reported results of a MIPS GTO survey of the
Pleiades, and identified nine cluster members that appear to
have 24 \mum\ excesses due to circumstellar disks.   However,
it is possible that in a few cases these apparent excesses could
be due instead to a knot of the passing interstellar dust
impacting the cluster member, or that the 24 \mum\ excess could
be flux from a background galaxy projected onto the line of
sight to the Pleiades member.   Careful analysis of the IRAC
images of these cluster members may help confirm that the MIPS
excesses are evidence for debris disks rather than the other
possible explanations.  

Six of the Pleiades members with probable 24 \mum\ excesses are
included in the region mapped with IRAC.  However, only four of
them have data at 8 \mum\ -- the other two fall near the edge of
the mapped region and only have data at 3.6 and 5.8 \mum.   
None of the six stars appear to have significant local nebular
dust from visual inspection of the IRAC mosaic images.   Also,
none of them appear problematic in Figure \ref{fig:dusty1}.  
For a slightly more quantitative analysis of possible nebular
contamination, we also constructed aperture growth curves for
the six stars, and compared them to other Pleiades members.  All
but one of the six show aperture growth curves that are normal
and consistent with the expected IRAC PSF.  The one exception is
HII 489, which has a slight excess at large aperture sizes as is
illustrated in Figure \ref{fig:ap_grow2}.   Because HII 489 only
has a small 24 \mum\ excess, it is possible that the 24 \mum\
excess is due to a local knot of the interstellar cloud material
and is not due to a debris disk.   For the other five 24 \mum\
excess stars we find no such problem, and we conclude that their
24 \mum\ excesses are indeed best explained as due to debris
disks.

\section{Summary and Conclusions}

We have collated the primary membership catalogs for the Pleiades to
produce the first catalog of the cluster extending from its highest
mass members to the substellar limit.  At the bright end, we expect
this catalog to be essentially complete and with few or no non-member
contaminants.   At the faint end, the data establishing membership are
much sparser, and we expect a significant number of objects will be
non-members.   We hope that the creation of this catalog will spur
efforts to obtain accurate radial velocities and proper motions for
the faint candidate members in order to eventually provide a
well-vetted membership catalog for the stellar members of the
Pleiades.   Towards that end, it would be useful to update the current
catalog with other data -- such as radial velocities, lithium
equivalent widths, x-ray fluxes, H$\alpha$ equivalent widths, etc.\ -- 
which could be used to help accurately establish membership for the
low mass cluster candidates.  It is also possible to make more use of
``negative information" present in the proper motion catalogs. That
is, if a member from one catalog is not included in another study but
does fall within its areal and luminosity coverage, that suggests that
it likely failed the membership criteria of the second study.   For a
few individual stars, we have done this type of comparison, but a
systematic analysis of the proper motion catalogs should be
conducted.  We intend to undertake these tasks,  and plan to establish
a website where these data would be hosted.

We have used the new Pleiades member catalog to define the
single-star locus at 100 Myr for $BVI_c$\ks\ and the four IRAC
bands.  These curves can be used as empirical calibration curves
when attempting to identify members of less well-studied, more
distant clusters of similar age. We compared the Pleiades
photometry to theoretical isochrones from \citet{siess00} and
\citet{baraffe98}.  The \citet{siess00} isochrones are not, in
detail, a good fit to the Pleiades photometry, particularly for
low mass stars.   The \citet{baraffe98} 100 Myr isochrone does
fit the Pleiades photometry very well in the $I$ vs.\ $I-K$ plane.

We have identified 31 new substellar candidate members of the
Pleiades using our combined seven-band infrared photometry, and
have shown that the majority of these objects appear to share the
Pleiades proper motion.   We believe that most of the objects
that may be contaminating our list of candidate brown dwarfs are
likely to be unresolved galaxies, and therefore low resolution
spectroscopy should be able to provide a good criterion for
culling our list of non-members.

The IRAC images, particularly the 8 \mum\ mosaic, provide vivid
evidence of the strong interaction of the Pleiades stars and the
interstellar cloud that is passing through the Pleiades.  Our
data are supportive of the model proposed by \citet{herbig01}
whereby the passing cloud is part of the Taurus cloud complex
and hence is encountering the Pleiades from the SSE direction.  
\citet{white93} had proposed a model whereby the cloud was encountering
the Pleiades from the west and used this to explain features in
the IRAS 60 and 100 $\mu$m images of the region as the wake of
the Pleiades moving through the cloud.  Our data appear to not be
supportive of that hypothesis, and therefore leaves the apparent
structure in the IRAS maps as unexplained.

\acknowledgments

Most of the support for this work was provided by the Jet
Propulsion Laboratory, California Institute of Technology, under
NASA contract 1407.   This research has made use of NASA's
Astrophysics Data System (ADS) Abstract Service, and of the
SIMBAD database, operated at CDS, Strasbourg, France.  This
research has made use of data products from the Two Micron
All-Sky Survey (2MASS), which is a joint project of the
University of Massachusetts and the Infrared Processing and
Analysis Center, funded by the National Aeronautics and Space
Administration and the National Science Foundation.  These data
were served by the NASA/IPAC Infrared Science Archive, which is
operated by the Jet Propulsion Laboratory, California Institute
of Technology, under contract with the National Aeronautics and
Space Administration.  The research described in this paper was
partially carried out at the Jet Propulsion Laboratory,
California Institute of Technology, under contract with the
National Aeronautics and Space Administration.  

\appendix

\section{APPENDIX}

\subsection{Membership Catalogs}

Membership lists of the Pleiades date back to antiquity if one
includes historical and literary references to the Seven Sisters
(Alcyone, Maia, Merope, Electra, Taygeta, Asterope and Celeno)
and their parents (Atlas and Pleione).  The first paper
discussing relative proper motions of a large sample of
stars in the Pleiades (based on visual observations)
was published by
\citet{pritchard84}.   The best of the early proper motion
surveys of the Pleiades derived from photographic plate
astrometry was that by \citet{trumpler21}, based on plates
obtained at Yerkes and Lick observatories. The candidate members
from that survey were presented in two tables, with the first
being devoted to candidate members within about one degree from
the cluster center (operationally, within one degree from
Alcyone) and the second table being devoted to candidates
further than one degree from the cluster center.  Most of the
latter stars  were denoted  by Trumpler by an S or R, followed
by an identification number.  We use Tr to designate the
Trumpler stars (hence Trnnn for a star from the 1st table and
the small number of stars in the second table without an ``S" or
an ``R", and TrSnnn or TrRnnn for the other stars).  For the
central region, Trumpler's catalog extends to $V \sim$ 13, while
the outer region catalog includes stars only to about $V \sim$
9.

The most heavily referenced proper motion catalog of the
Pleiades is that provided by \citet{hertzsprung47}.   That paper
makes reference to two separate catalogs: a photometric catalog
of the Pleiades published by Hertzsprung in 1923
\citep{hertzsprung23}, whose members are commonly referred to by
HI numbers, and the new proper motion catalog from the 1947
paper, commonly referenced as the HII catalog.   While both HI
and HII numbers have been used in subsequent observational
papers, it is the HII identification numbers that predominate. 
That catalog -- derived from Carte du Ciel blue-sensitive plates
from 14 observatories -- includes stars in the central
2$\times$2 degree region of the cluster, and has a faint limit
of about  $V$ = 15.5.   Johnson system $BVI$ photometry is
provided for most of the proposed Hertzsprung members in
\citet{jomi58} and \citet{iriarte67}.  Additional Johnson $B$ and
$V$ photometry plus Kron $I$ photometry for a fairly large number of
the Hertzsprung members can be found in \citet{stauffer80},
\citet{stauffer82}, and \citet{stauffer84}.   Other Johnson $BV$
photometry for a scattering of stars can be found in
\citet{jones73}, \citet{robinson74},  \citet{messina01}. 
Spectroscopic confirmation, primarily via radial velocities,
that these are indeed Pleiades members has been provided in 
\citet{soderblom93,queloz98} and \citet{mermilliod97}.

Two other proper motion surveys provide relatively bright
candidate members relatively far from the cluster center:
\citet{artyukhina70} and  \citet{vanlee86}.  Stars from the
Artyukhina catalog are designated as AK followed by the region
from which the star was identified followed by an identification
number.   The new members provided in the van Leeuwen paper were
taken from an otherwise unpublished proper motion study by Pels,
where the first 118 stars were considered probable members and
the remaining 75 stars were considered possible members.  Van
Leeuwen categorized a number of the Pels stars as non-members
based on the Walraven photometry they obtained, and we adopt
those findings.   Radial velocities for stars in these two
catalogs have been obtained by \citet{rosvick92},
\citet{mermilliod97}, and \citet{queloz98}, 
and those authors identified a list of
the candidate members that they considered confirmed by the high
resolution spectroscopy.  For these outlying candidate members,
to be included in Table 2 we require that the star be a radial
velocity member from one of the above three surveys, or be
indicated as having ``no dip" in the Coravel cross-correlation
(indicating rapid rotation, which at least for the later type
stars is suggestive of membership).
Geneva photometry of the Artyukhina stars considered as likely
members was provided by \citet{mermilliod97}.  The magnitude
limit of these surveys was not well-defined, but most of the
Artyukhina and Pels stars are brighter than $V$=13.

\citet{jones73} provided proper motion membership probabilities
for a large sample of proposed Pleiades members, and for a set
of faint, red stars towards the Pleiades.  A few star identification
names from the sources considered by Jones appear in Table 2,
including MT \citep{mccarthy64}, VM \citep{vanmaanen46},
ALR \citep{ahmed65}, and J \citep{jones73}.

The chronologically next significant source of new Pleiades
candidate members was the flare star survey of the Pleiades
conducted at several observatories in the  1960s, and
summarized in \citet{haro82}, hereafter HCG.    The logic
behind these surveys was that even at 100 Myr, late type dwarfs
have relatively frequent and relatively high luminosity flares
(as demonstrated by \citet{jomi58} having detected two flares
during their photometric observations of the Pleiades), and
therefore wide area, rapid cadence imaging of the Pleiades at
blue wavelengths should be capable of identifying low mass
cluster members.  However, such surveys also will detect
relatively young field dwarfs, and therefore it is best to
combine the flare  star surveys with proper motions.   Dedicated
proper motion surveys of the HCG flare stars were conducted by 
\citet{jones81} and \citet{stauffer91}, with the latter also
providing photographic $VI$ photometry (Kron system).  
Photoelectric photometry for some of the HCG stars have been
reported in \citet{stauffer82}, \citet{stauffer84},
\citet{stauffer87}, and \citet{prosser91}.  High resolution
spectroscopy of many of the HCG stars is reported in
\citet{stauffer84}, \citet{stauffer87} and \citet{terndrup00}. 
Because a number of the papers providing additional
observational data for the flare stars were obtained prior to
1982, we also include in Table 2 the original flare star names
which were derived from the observatory where the initial flare
was detected.  Those names are of the form an initial letter
indicating the observatory -- A (Asiago),  B (Byurakan), K
(Konkoly), T (Tonantzintla) -- followed by an identification
number.

\citet{stauffer91} conducted two proper motion surveys of the
Pleiades over an approximately 4$\times$4 degree region of the
cluster based on plates obtained with the Lick
20$^{\prime\prime}$ astrographic telescope.   The first survey
was  essentially unbiased, except for the requirement that the
stars fall approximately in the region of the $V$ vs.\ $V-I$
color-magnitude diagram where Pleiades members should  lie. 
Candidate members from this survey are designated by SK numbers.
The second survey was a proper motion survey of the HCG stars. 
Photographic $VI$ photometry of all the stars was provided as well
as proper motion membership probabilities.   Photoelectric
photometry for some of the candidate members was obtained as
detailed above in the section on the HCG catalog stars.  The
faint limit of these surveys is about $V$=18.

\citet{hambly91} provided a significantly deeper, somewhat wider
area proper motion survey, with the faintest members having V
$\simeq$ 20 and the total area covered being of order 25 square
degrees.  The survey utilized red sensitive plates from the
Palomar and UK Schmidt telescopes.  Due to  incomplete coverage
at one epoch, there is a vertical swath slightly east of the
cluster center where no membership information is available.  
Stars from this survey are designated by their HHJ numbers . 
\citet{hambly93} provide $RI$ photographic photometry on a natural
system for all of their candidate members, plus photoelectric
Cousins $RI$ photometry for a small number of stars and $JHK$
photometry for a larger sample.  Some spectroscopy to confirm
membership has been reported in \citet{stauffer94},
\citet{stauffer95}, \citet{oppenheimer97}, \citet{stauffer98},
and \citet{steele95}, though for most of the HHJ stars there is
no spectroscopic membership confirmation.  

\citet{pinfield00} provide the deepest wide-field proper motion
survey of the Pleiades.
That survey combines CCD imaging of six square degrees
of the Pleiades obtained with the Burrell Schmidt telescope (as
five separate, non-overlapping fields near but outside the
cluster center) with deep photographic plates which provide the
1st epoch positions.   Candidate members are designated by BPL
numbers (for Burrell Pleiades), with the faintest stars having
$I\simeq$ 19.5, corresponding to $V >$ 23.  Only the stars
brighter than about $I$= 17 have sufficiently accurate proper
motions to use to identify Pleiades members.  Fainter than $I$=
17, the primary selection criteria are that the star fall in an
appropriate place in both an $I$ vs.\ $I-Z$ and an $I$ vs.\
$I-K$ CMD. 

\citet{adams01} combined the 2MASS and digitized POSS databases to produce
a very wide area proper motion survey of the Pleiades.  By
design, that survey was very inclusive - covering the entire
physical area of the cluster and extending to the hydrogen burning
mass limit.   However, it was also very ``contaminated", with
many suspected non-members.   The catalog  of possible members
was not published.  We have therefore not included stars from
this study in Table 2; we have used the proper motion data from
\citet{adams01} to help decide cases where a given star has
ambiguous membership data from the other surveys.  

\citet{deacon04} provided another deep and very wide area proper
motion survey of the Pleiades.
The survey covers a circular area of approximately five
degrees radius to $R \sim$\ 20, or $V \sim$ 22.   Candidate
members are designated by DH.  \citet{deacon04} also provide
membership probabilities based on proper motions for many
candidate cluster members from previous surveys.   For stars
where \citet{deacon04} derive P $<$ 0.1 and where we have no
other proper motion information or where another proper motion
survey also finds low membership probability, we exclude the
star from our catalog.   For cases where two of our proper
motion catalogs differ significantly in their membership
assessment, with one survey indicating the star is a probable
member, we retain the star in the catalog as the conservative
choice.  Examples of the latter where \citet{deacon04} derive P
$<$ 0.1 include HII 1553, HII 2147, HII 2278 and HII 2665 -- all of
which we retain in our catalog because other surveys indicate
these are high probability Pleiades members.  

\subsection{Photometry}

Photometry for stars in open cluster catalogs can be used to
help confirm cluster membership and to help constrain physical
properties of those stars or of the cluster.    For a variety of
reasons, photometry of stars in the Pleiades has been obtained
in a panoply of different photometric systems.  For our own
goals, which are to use the photometry to help verify
membership and to define the Pleiades single-star locus in color
magnitude diagrams, we have attempted to convert photometry in
several of these systems to a common system (Johnson $BV$ and
Cousins $I$).   We detail below the sources of the  photometry and
the conversions we have employed.

Photoelectric photometry of Pleiades members dates back to at
least 1921 \citep{cummings21}.  However, as far as we are aware
the first ``modern" photoelectric photometry for the Pleiades,
using a potassium hydride photoelectric cell, 
is that of \citet{calder37}.  \citet{eggen50}
provided photoelectric photometry using a 1P21 phototube
(but calibrated to a no-longer-used  photographic system) for
most of the known Pleiades members within one degree of the
cluster center and with magnitudes $<$ 11.  The first phototube
photometry of Pleiades stars calibrated more-or-less to the
modern UBV system was provided by \citet{jomo51}. An update of
that paper, and the oldest photometry included here was reported
in \citet{jomi58}, which provided $UBV$ Johnson system photometry
for a large sample of HII and Trumpler candidate Pleiades
members.   \citet{iriarte67} later reported Johnson system $V-I$
colors for most of these stars.   We have converted Iriarte's
$V-I$ photometry to  estimated Cousins $V-I$ colors using a formula
from \citet{bessell79}:
\begin{equation}
V - I ({\rm Cousins}) = 0.778 \times V - I ({\rm Johnson}).
\end{equation}
$BVRI$ photometry for most of the Hertzsprung members fainter than
$V$= 10 has been published by \citet{stauffer80},
\citet{stauffer82}, \citet{stauffer84}, and
\citet{stauffer87}.   The $BV$ photometry is Johnson system,
whereas the $RI$ photometry is on the Kron system.   The Kron
$V-I$
colors were converted to Cousins $V-I$ using a transformation
provided by \citet{bessell87}:
\begin{equation}
V - I ({\rm Cousins}) = 0.227 + 0.9567(V-I)_k +0.0128(V-I)_k^2 - 
0.0053(V-I)_k^3
\end{equation}
Other Kron system $V-I$ colors have been published for Pleiades
candidates in \citet{stauffer91} (photographic photometry) and
in \citet{prosser91}. These Kron-system colors have also been
converted to Cousins $V-I$ using the above formula.

Johnson/Cousins $UBVR$ photometry for a set of low mass Pleiades
members was provided by \citet{landolt79}.  We only use the $BV$
magnitudes from that study. Additional Johnson system $UBV$
photometry for small numbers of stars is provided in
\citet{robinson74}, \citet{messina01} and \citet{jones73}.

\citet{vanlee87} provided Walraven $VBLUW$ photometry for nearly
all of the Hertzsprung members brighter than $V \sim$ 13.5 and
for the Pels candidate members.  Van Leeuwen provided an
estimated Johnson $V$ derived from the Walraven $V$ in his tables. 
We have transformed the Walraven $V-B$ color into an estimate of
Johnson $B-V$ using a formula from \citet{rosvick92}:
\begin{equation}
B - V ({\rm Johnson}) = 2.571(V-B) -1.02(V-B)^2 +0.5(V-B)^3 -0.01
\end{equation}
\citet{hambly93} provided photographic $VRI$ photometry for all
of the HHJ candidate members, and $VRI$ Cousins photoelectric
photometry for a small fraction of those stars.  We took all of
the HHJ stars with  photographic photometry for which we also
have photoelectric $VI$ photometry on the Cousins system, and
plotted $V$(Cousins) vs.\ $V$(HHJ) and $I$(Cousins) vs.\
$I$(HHJ).  While there is some evidence for slight systematic
departures of the HHJ photographic photometry from the Cousins
system, those departures are relatively small and we have chosen
simply to retain the HHJ values and treat them as Cousins
system.  

\citet{pinfield00} reported their $I$ magnitudes in an
instrumental system which they designated as $I_{kp}$.  We
identified all BPL candidate members for which we had
photoelectric Cousins I estimates, and plotted $I_{kp}$ vs.\
\ic.  Figure~\ref{fig:ikpic} shows this correlation, and the
piecewise linear fit we have made to convert from $I_{kp}$ to
\ic.  Our catalog lists these converted \ic\ measures for the
BPL stars for which we have no other photoelectric $I$ estimates.

\citet{deacon04} derived $RI$ photometry from the scans of their
plates, and calibrated that photometry by reference to published
photometry from the literature.   When we plotted their the difference
between their $I$ band photometry and literature values (where
available), we discovered a significant dependence on right
ascension.  Unfortunately, because the DH survey extended over larger
spatial scales than the calibrating photometry, we could not derive a
correction which we could apply to all the DH stars.   We therefore
developed the following  indirect scheme.   We used the stars for
which we have estimated \ic\ magnitudes (from photoelectric
photometry) to define the relation between $J$ and (\ic$ - J$) for
Pleiades members.  For each DH star, we combined that relation and the
2MASS $J$ magnitude to yield a predicted \ic.   Figure
\ref{fig:dh_ra} shows a plot of the difference of this predicted \ic\
and $I$(DH) with right ascension.    The solid line shows the relation
we adopt.    Figure \ref{fig:dh_icorr} shows the relation between the
corrected $I$(DH) values and Table 2 \ic\ measures from photoelectric
sources.  There is still a significant amount of scatter but the
corrected $I$(DH) photometry appears to be accurately calibrated to the
Cousins system.

In a very few cases (specifically, just five stars), we 
provide an estimate of I$_c$\ based on data
from a wide-area CCD survey of Taurus obtained with the Quest-2 camera
on the Palomar 48 inch Samuel Oschin telescope \citep{slesnick06}.  That
survey calibrated their photometry to the Sloan i system, and we have
converted the Sloan i magnitudes to I$_c$.  We intend to make more
complete use of the Quest-2 data in a subsequent paper.

When we have multiple sources of photometry for a given star, we
consider how to combine them.   In most cases, if we have
photoelectric data, that is given preference.   However, if we have
photographic $V$ and $I$, and only a photoelectric measurement for
$I$, we do not replace the photographic $I$ with the photoelectric
value because these stars are variable and the photographic
measurements are at least in some cases from nearly simultaneous
exposures.   Where we have multiple sources for photoelectric
photometry, and no strong reason to favor one measurement or set of
measurements over another, we have averaged the photometry for a given
star.   In most cases, where we have multiple photometry the
individual measurements agree reasonably well but with the caveat that
the Pleiades low mass stars are in many cases heavily spotted and
``active" chromospherically, and hence are photometrically variable. 
In a few cases, even given the expectation that spots and other
phenomena may affect the photometry, there seems to be more
discrepancy between reported $V$ magnitudes than we expect.  We note
two such cases here.  We suspect these results indicate that at least
some of the Pleiades low mass stars have long-term photometric
variability larger than their short period (rotational) modulation.

HII 882 has at least four presumably accurate $V$ magnitude
measurements reported in the literature.  Those measures are:  
$V$=12.66 \citet{jomi58}; $V$=12.95 \citet{stauffer82};  $V$=12.898
\citet{vanlee86};  and $V$=12.62 \citet{messina01}.

HII 345 has at least three presumably accurate $V$ magnitude
measurements. Those measurements are:  $V$=11.65
\citet{landolt79}; $V$=11.73 \citet{vanlee86}; $V$=11.43
\citet{messina01}.

At the bottom of Table 2, we provide a key to the source(s) of the
optical photometry provided in the table.  

\acknowledgments

This research made use of the SIMBAD database operated at CDS,
Strasbourg, France, and also of the NED and NStED databases
operated at IPAC, Pasadena, USA.  A large amount of data for the Pleiades
(and other open clusters) can also be found at the open cluster
database WEBDA (http://www.univie.ac.at/webda/), operated in
Vienna by Ernst Paunzen.



\begin{deluxetable}{lcccc}
\tablecaption{Pleiades Membership Surveys used as Sources}
\tablewidth{0pt}
\tablehead{
\colhead{Reference} &  \colhead{Area Covered} &
\colhead{Magnitude Range } & \colhead{Number Candidates} & 
\colhead{Name}   \\
           & \colhead{Sq. Deg.}& 
           \colhead{(and band)}&  &\colhead{Prefix}}
\startdata
Trumpler (1921) &    3     &  2.5$<B<$14.5 &   174    &   Tr \\
Trumpler (1921)\tablenotemark{a} &   24     &  2.5$<B<$10   &    72    &   Tr \\
Hertzsprung (1947) & 4     &  2.5$<V<$15.5 &   247    &   HII \\ 
Artyukhina (1969) & 60     &  2.5$<B<$12.5 &  $\sim$200 &  AK \\
Haro \etal\ (1982) &       20     &   11$<V<$17.5  &   519   &   HCG  \\
van Leeuwen \etal\ (1986) & 80     &  2.5$<B<$13   &   193   &   PELS  \\ 
Stauffer \etal\ (1991) &   16     &   14$<V<$18     &  225   &   SK   \\ 
Hambly \etal\ (1993)   &   23     &   10$<I<$17.5   & 440    &   HHJ  \\
Pinfield \etal\ (2000) &    6     & 13.5$<I<$19.5 & 339    &   BPL  \\
Adams \etal\ (2001)    &   300    &    8$<Ks<$14.5 & 1200  &  ...  \\
Deacon \& Hambly (2004)   &   75     &   10$<R<$19 &     916    &   DH  \\
\enddata 
\tablenotetext{a}{The Trumpler paper is listed twice
because there are two membership surveys included in that paper, with
differing spatial coverages and different limiting magnitudes.}
\end{deluxetable}

\include{table2}

\include{table3}

\include{table4}

\include{table5}


\clearpage
\newpage

\pagestyle{myheadings}
\pagenumbering{arabic}
\setcounter{page}{1}


\clearpage
\newpage

\begin{figure} %
    \caption{ Spatial coverage of the six times deeper ``2MASS 6x" 
       observations of the Pleiades.  The
       2MASS survey region is approximately centered on Alcyone, the most massive
       member of the Pleiades.  The trapezoidal box roughly indicates the
       region covered with the shallow IRAC survey of the cluster core.  The star
       symbols correspond to the brightest B star members of the cluster.  The red
       points are the location of objects in the 2MASS 6x Point Source Catalog.
    \label{fig:2maspatial}
    }
\end{figure}


\clearpage
\newpage

\begin{figure} %
    \caption{
       Color-magnitude diagram for the Pleiades derived from the 2MASS 6x
       observations.  The red dots correspond to objects identified as unresolved,
       whereas the green dots correspond to extended sources (primarily background
       galaxies).   The lack of green dots fainter than K = 16 is indicative
       that there is too few photons to identify sources as extended - the
       extragalactic population presumably increases to fainter magnitudes.
    \label{fig:2macmd}
    }
\end{figure}


\clearpage
\newpage

\begin{figure} %
    \caption{
       As for Figure 2, except in this case the axes are $J-H$
       and $H-K_s$.
       The extragalactic objects are very red in both colors.
    \label{fig:2maccd}
    }
\end{figure}


\clearpage
\newpage

\begin{figure} %
   \caption{
        {\em FIGURE REMOVED TO FIT WITHIN ASTRO-PH FILESIZE GUIDELINES. See
        http://spider.ipac.caltech.edu/staff/stauffer/pleiades07/ for
        full-res version.}
       Two-color (4.5 and 8.0 micron) mosaic of the central square degree
       of the Pleiades from the IRAC survey.  North is approximately vertical,
       and East is approximately to the left.  The bright star nearest the
       center is Alcyone; the bright star at the left of the mosaic is
       Atlas; and the bright star at the right of the mosaic is Electra.
   \label{fig:pleIRAC}
   }
\end{figure}


\clearpage
\newpage

\begin{figure} %
   \caption{
        Finding chart corresponding approximately to the region
        imaged with IRAC. The large,
        five-pointed stars are all of the Pleiades members
        brighter than $V$= 5.5.  The small open circles
        correspond to other cluster members.  Several stars with
        8 \mum\ excesses are labelled by their HII numbers, and
        are discussed further in Section 6.  The short lines
        through several of the stars indicate the size and
        position angle of the residual optical polarization
        (after subtraction of a constant foreground component),
        as provided in Figure 6 of \citet{breger86}.
   \label{fig:cartoon}
   }
\end{figure}


\clearpage
\newpage

\begin{figure} %
   \caption{
       Comparison of aperture photometry for Pleiades members derived from
       the IRAC 3.6 \mum\ mosaic using the Spitzer APEX package and the
       IRAF implementation of DAOPHOT.
   \label{fig:plecomp1}
   }
\end{figure}


\clearpage
\newpage

\begin{figure} %
   \caption{
       Difference between aperture photometry for Pleiades members for IRAC
       channels 1 and 2.  The [3.6]$-$[4.5] color begins to depart from
       essentially zero at magnitudes $\sim$10.5, corresponding
       approximately to spectral type M0 in the Pleiades. 
   \label{fig:plecomp2}
   }
\end{figure}


\clearpage
\newpage

\begin{figure} %
   \caption{
       $K_s$ vs.\ $K_s$ $-$[4.5] CMD for Pleiades candidate members, illustrating
       why we have excluded HII 1695 from the final catalog of cluster members.
       The ``X" symbol marks the location of HII 1695 in this diagram.
   \label{fig:ple1695}
   }
\end{figure}


\clearpage
\newpage

\begin{figure} %
   \caption{
       Spatial plot of the candidate Pleiades members from Table 2.  The
       large star symbols are members brighter than \ks= 6; the open
       circles are stars with 6 $<$ \ks\ $<$ 9; and the dots are candidate
       members fainter than \ks= 9.  The solid line is parallel to
       the galactic plane.
   \label{fig:plespatial2}
   }
\end{figure}


\clearpage
\newpage

\begin{figure} %
   \caption{
       The cumulative radial density profiles for Pleiades members in several
       magnitude ranges:  heavy, long dash -- \ks\ $<$ 6; dots
       -- 6 $<$ \ks\ $<$ 9; short dash -- 9 $<$ \ks\ $<$ 12;
       light, long dash -- \ks\ $>$ 12.
   \label{fig:ple_segreg}
   }
\end{figure}


\clearpage
\newpage

\begin{figure} %
   \caption{
       $V$ vs.\ $(V-I)_c$ CMD for Pleiades members with photoelectric photometry.
       The solid curve is the ``by eye" fit to the single-star locus for
       Pleiades members.
   \label{fig:cmd_vmi}
   }
\end{figure}


\clearpage
\newpage

\begin{figure} %
   \caption{
       $K_s$ vs.\ $K_s-[3.6]$ CMD for Pleiades candidate members
       from Table 2 (dots) and from deeper imaging of a set of
       Pleiades VLM and brown dwarf candidate members from
       \citet{lowrance07} (squares).   The solid curve is the
       single-star locus from Table 3.
   \label{fig:cmd_km1}
   }
\end{figure}


\clearpage
\newpage

\begin{figure} %
   \caption{
       $V$ vs.\ $(V-I)_c$ CMD for Pleiades candidate members from Table 2 for
       which we have photoelectric photometry,
       compared to theoretical isochrones from \citet{siess00} (left) and
       from \citet{baraffe98} (right).  For the left panel, the curves
       correspond to 10, 50, 100 Myr and a ZAMS; the right panel includes
       curves for 50, 100 Myr and a ZAMS.
   \label{fig:super_vmi}
   }
\end{figure}


\clearpage
\newpage

\begin{figure} %
   \caption{
       $K$ vs.\ $(I-K)$ CMD for Pleiades candidate members from Table 2,
       compared to theoretical isochrones from \citet{siess00} (left) and
       from \citet{baraffe98} (right).  The curves
       correspond to  50 Myr, 100 Myr and a ZAMS. 
   \label{fig:super_kik}
   }
\end{figure}


\clearpage
\newpage

\begin{figure} %
   \caption{
       \ks\ vs.\ \ks$-$[3.6] CMD for the objects in the central
       one square degree of the Pleiades, combining data from the IRAC
       shallow survey and 2MASS.  The symbols are defined within the
       figure (and see text for details).   The dashed-line box indicates
       the region within which we have searched for new candidate Pleiades
       VLM and substellar members.  The solid curve is a DUSTY 100 Myr
       isochrone from \citet{chabrier00}, for masses from 0.1 to 0.03 \msun.
   \label{fig:cmd3dot6}
   }
\end{figure}


\clearpage
\newpage

\begin{figure} %
   \caption{
       Proper motion vector point diagrams (VPDs) for various
       stellar samples in the central one degree field, derived
       from combining the IRAC and 2MASS 6x observations.  Top
       left: VPD comparing all objects in the field (small black
       dots) to Pleiades members  with 11 $<$ \ks\ $<$ 14 (large
       blue dots). Top right: same, except the blue dots are the
       new candidate VLM and  substellar Pleiades members.
       Bottom left: same, except the blue dots are a nearby, low
       mass field star sample from a box just blueward of  the
       trapezoidal region in \ref{fig:cmd3dot6}. Bottom
       right: VPD just showing a second, distant field star
       sample as described in the text.
   \label{fig:super_propmo}
   }
\end{figure}


\clearpage
\newpage

\begin{figure} %
   \caption{
       Same as Fig.~\ref{fig:cmd3dot6}, except that the new candidate VLM and
       substellar objects from Table 4 are now indicated as small, red squares.
   \label{fig:cmd3dot6memb}
   }
\end{figure}


\clearpage
\newpage

\begin{figure} %
   \caption{
       Two plots intended to isolate Pleiades members with
       excess and/or extended 8 \mum\ emission.   The plot with
       [3.6]$-$[8.0] micron colors shows data from Table 3 (and
       hence is for aperture sizes of 3 pixel and 2 pixel
       radius, respectively).  The increased  vertical spread in
       the  plots at faint magnitudes is simply due to
       decreasing signal to noise at 8 \mum.  The numbers
       labelling stars with excesses are the HII identification
       numbers for those stars.
   \label{fig:dusty1}
   }
\end{figure}


\clearpage
\newpage

\begin{figure} %
   \caption{
       Postage stamp images extracted from individual, 8 \mum\ BCDs
       for the stars with extended 8 \mum\ emission, from which we
       have subtracted an empirical PSF.  Clockwise from the upper
       left, the stars shown are HII1234, HII859, Merope and HII652.
       The five-pointed star indicates the astrometric
       position of the star (often superposed on a few black pixels
       where the 8 \mum\ image was saturated.  The circle in the
       Merope image is centered on the location of IC349 and has
       diameter about 25" (the size of IC349 in the optical is of
       order 10" $x$ 10").
   \label{fig:psfsub}
   }
\end{figure}


\clearpage
\newpage

\begin{figure} %
   \caption{
       Aperture growth curves from the 8 \mum\ mosaic for stars with
       24 \mum\ excesses from \citet{gorlova06} and for a set of
       control objects (dashed curves).   All of the objects have been
       scaled to common zero-point magnitudes for 9 pixel apertures,
       with the 24 \mum\ excess stars offset from the control objects
       by 0.1 mag.  The three \citet{gorlova06} stars with no excess at
       8 \mum\ are HII 996, HII 1284 and HII 2195.  The \citet{gorlova06}
       star with a slight excess at 8 \mum\ is HII 489.
   \label{fig:ap_grow2}
   }
\end{figure}


\clearpage
\newpage

\begin{figure} %
   \caption{
       Calibration derived relating $I_{kp}$\ from \citet{pinfield00}
       and \ic.  The dots are stars for which we have both $I_{kp}$
       and \ic\ measurements (small dots: photographic \ic;
       large dots: photoelectric \ic), and the solid line indicates the 
       piecewise linear fit we use to convert the $I_{kp}$ values to
       \ic\ for stars for which we only have $I_{kp}$.
    \label{fig:ikpic}
    }
\end{figure}


\clearpage
\newpage

\begin{figure} %
   \caption{
       Difference between the predicted \ic\ and \citet{deacon04}
       $I$ magnitude as a function of right ascension for the DH stars.
       No obvious dependence is present versus declination.
    \label{fig:dh_ra}
    }
\end{figure}


\clearpage
\newpage

\begin{figure} %
   \caption{
       Comparison of the recalibrated DH $I$ photometry with estimates
       of \ic\ for stars in Table 2 with photoelectric data.
    \label{fig:dh_icorr}
    }
\end{figure}

\end{document}